# Multi-stable dissipative structures pinned to dual hot spots

Cheng Hou Tsang[1], Boris A. Malomed [2, 3], and Kwok Wing Chow[1,*]

[1]Department of Mechanical Engineering, University of Hong Kong, Pokfulam Road, Hong Kong

[2]Department of Physical Electronics, School of Electrical Engineering, Faculty of Engineering, Tel Aviv University, Tel Aviv 69978, Israel

[3]ICFO-Institut de Ciencies Fotoniques, Mediterranean Technology Park, 08860 Castelldefels (Barcelona), Spain (*temporary Sabbatical address*)

*Corresponding author. Fax: (852) 2858 5415. Email address: kwchow@ hku.hk

**Abstract**

   We analyze the formation of one-dimensional localized patterns in a nonlinear dissipative medium including a set of two narrow "hot spots" (HSs), which carry the linear gain, local potential, cubic self-interaction, and cubic loss, while the linear loss acts in the host medium. This system can be realized, as a spatial-domain one, in optics, and also in Bose-Einstein condensates of quasi-particles in solid-state settings. Recently, exact solutions were found for localized modes pinned to the single HS represented by the delta-function. The present paper reports analytical and numerical solutions for coexisting two- and multi-peak modes, which may be symmetric or antisymmetric with respect to the underlying HS pair. Stability of the modes is explored through simulations of their perturbed evolution. The sign of the cubic nonlinearity plays a crucial role: in the case of the self-focusing, only the fundamental symmetric and antisymmetric modes, with two local peaks tacked to the HSs, and no additional peaks between them, may be stable. In this case, all the higher-order multi-peak modes, being unstable, evolve into the fundamental ones. Stability regions for the fundamental modes are reported.  A more interesting situation is found in the case of the self-defocusing cubic nonlinearity, with the HS pair giving rise to a multi-stability, with up to eight coexisting stable multi-peak patterns, symmetric and antisymmetric ones. The system without the self-interaction, the nonlinearity being represented only by the local cubic loss, is investigated too. This case is similar to those with the self-focusing or defocusing nonlinearity, if the linear potential of the HS is, respectively, attractive or repulsive. An additional feature of the former setting is the coexistence of the stable fundamental modes with robust breathers.

## I. INTRODUCTION AND THE MODEL

One of the fundamental effects in photonics is self-trapping of spatial solitons in nonlinear waveguides [1-3]. This was demonstrated experimentally in media with the cubic (Kerr) [4], quadratic [5], photorefractive (saturable) [6] and nonlocal [7] nonlinearities, as well as in various lattice media, based on arrayed waveguides [1]. A relatively small longitudinal size of samples used in the experiments makes it possible to neglect the loss in the corresponding models [1-3]. On the other hand, it is crucially important to take the loss and compensating gain into account in the analysis of the light generation and transmission in laser cavities, where the nonlinear waveguide is a part of an optical loop, see original papers [8] and books [9]. The corresponding cavity models, which include the transverse diffraction of light, Kerr nonlinearity, gain, and the background loss, are based on complex Ginzburg-Landau (CGL) equations. In these models, the spatial dissipative solitons are supported by the simultaneous diffraction/self-focusing and loss/gain balance [9].

A condition necessary for the stability of dissipative solitons is the stability of the zero background, which rules out single-component equations including the linear gain (the stability of dissipative solitons, including those given by exact solutions [10], may be provided in a system of linearly coupled CGL equations, with the linear gain applied in one component (core), and the linear loss acting in the additional, stabilizing core [10-12]). Stable dissipative solitons may be generated by the CGL equation of the cubic-quintic (CQ) type, which includes linear and quintic loss terms and the cubic gain. This possibility was first proposed in Ref. [13] and then elaborated by means of diverse analytical and numerical methods [14].

In Refs. [15] and [16], it was proposed to sustain stable dissipative solitons within the framework of the most fundamental setting, based on the single CGL equation with the cubic nonlinearity, while the linear gain is applied within a "hot spot" (HS), i.e., a narrow stripe embedded into a lossy waveguide. The HS can be induced by a strongly concentrated density of dopant atoms which provide for the gain, or simply by tightly focusing the laser beam which pumps the cavity. In terms of the theoretical model, the local gain concentrated at the HS may be approximated by a delta-function, see Eq. (1) below. Dissipative solitons pinned to the HS are stable due to the balance between the power supply from the HS and dissipation caused by the bulk loss, which may be mutually adjusted with the help of the nonlinear self-focusing or defocusing [15, 16]. The $\delta$-functional approximation offers an advantage of finding the pinned-soliton solutions in an *exact* analytical form [15], combining the *ansatz* suggested by the *Pereira-Stenflo* dissipative soliton of the cubic complex CGL equation (which is always unstable by itself) [17, 10] and the boundary conditions imposed by the $\delta$-function. In this case, exact solutions are not generic ones, as they are available under an additional constraint imposed on parameters of the system. Another solvable version of the model was proposed in Ref. [16], where the nonlinearity (both self-focusing/defocusing and cubic-loss terms) was also assumed to be concentrated at the HS, so that the $\delta$-function multiplies the cubic terms too. In that model, analytical solutions for the pinned dissipative solitons are generic. In terms of the physical realization, the latter model implies embedding of a narrow stripe of a pumped nonlinear material into a linear lossy waveguide, or the situation where both the gain and nonlinearity are provided by the strong concentration of dopants.

The localized gain belongs to a class of models based on diverse landscapes of the spatially inhomogeneous amplification, which have been recently elaborated in diverse settings. In particular, spatially periodic and localized stationary modes and breathers were investigated in the framework of the nonlinear-Schrödinger equation with parabolic [18], periodic [19], and double-well [20] complex potentials, whose imaginary part determines the spatially modulated gain/loss term. Related to the latter setting, is the analysis of symmetric, antisymmetric and asymmetric trapped states in the landscapes with two [21] or several [22] amplification channels. Also investigated were dissipative surface solitons pinned to the interface between uniform and periodic media [23], and vortices circulating along a two-dimensional ring to which the linear gain is applied [24]. It is also relevant to mention that related landscapes of local losses may be used for shaping various matter-wave patterns in atomic Bose-Einstein condensates [25].

A similar model was proposed earlier in Ref. [26], with the objective to trap a gap soliton in a fiber Bragg grating by means of the HS compensating the background loss. Related models were also proposed for other types of local defects in lasing media [27].

Another ramification is the analysis of patterns supported by a periodically distributed [28] or localized [29] injection, i.e., direct pump, instead of the local amplification. In addition to the description of laser cavities, the schemes with the direct pump are relevant as models of BEC of quasiparticles in solid-state media, such as magnons pumped by a microwave transducer [29].

Obviously interesting is a possibility to find *exact solutions* for localized patterns. As briefly mentioned in Ref. [16], in the model with both the gain and nonlinearity concentrated at HSs, it may be possible to construct analytical solutions for settings with a symmetric pair of the spots, in addition to the simplest localized modes supported by the single HS. The objective of the present work is to study symmetric and antisymmetric patterns supported by the paired HSs, which are represented by analytical solutions, or by related numerical ones [with the ideal $\delta$-functions replaced by their regularized counterparts, see Eq. (2) below]. The stability of the localized patterns is investigated by means of systematic direct simulations.

The analysis presented below produces results which strongly depend on the sign of the nonlinearity. In the case of the localized self-focusing, stable modes essentially amount to straightforward symmetric or antisymmetric double-peak superpositions of their counterparts supported by the single HSs. Essentially novel results are reported in the model with the self-defocusing nonlinearity, as well as in the case when the localized nonlinearity is represented solely by the cubic loss. These results include various stationary multi-peak patterns and chaotic or quasi-regular localized breathers, the coexistence of which gives rise to multi-stability. On the other hand, the settings considered in this work do not support nontrivial asymmetric modes, with respect to the underlying dual-HS structures. In that respect, our results are essentially different from those recently reported in the model with a smooth spatial distribution of the linear gain, where asymmetric modes emerge even in configurations with a single spatial maximum of the gain [22].

According to what was said above, the model considered in this paper is based on the CGL equation for the complex wave amplitude, $u(x,z)$, with the uniformly distributed linear loss, accounted for by coefficient $\gamma \geq 0$, while the gain and nonlinearity are concentrated at two HSs, which are set at points $x = \pm L$:

$$\frac{\partial u}{\partial z} = \frac{i}{2}\frac{\partial^2 u}{\partial x^2} - \gamma u + \left[(\Gamma_1 + i\Gamma_2) + (iB - E)|u|^2\right]\left[\delta(x+L) + \delta(x-L)\right]u. \qquad (1)$$

Equation (1) is written in the notation adjusted for the guided-wave optics, with $z$ and $x$ being the propagation distance and transverse coordinate, respectively. The scales of these variables are fixed, in what follows below, by setting $\gamma = L = 1$. Further, coefficients $\Gamma_1 > 0$ and $E \geq 0$ represent, severally, the gain and nonlinear loss concentrated at the HSs. Coefficient $\Gamma_2$ accounts for the possibility that the local refractive index may be altered inside the HS, thus inducing a local attractive ($\Gamma_2 > 0$) or repulsive ($\Gamma_2 < 0$) linear potential at each HS. Finally, the Kerr nonlinearity localized at the HSs is represented by coefficient $B > 0$ or $B < 0$, in the cases of the self-focusing and self-defocusing, respectively. If the local nonlinearity is induced by dopants, the sign of the nonlinearity is controlled by detuning of the double frequency of the carrier electromagnetic wave from the frequency of the intrinsic transition in the dopant atoms. The imaginary part of the Kerr coefficient, i.e., $E$ in Eq. (1), represents two-photon absorption, which is a well-known property of many optical materials; in particular, it may be enhanced in a semiconductor waveguide, when the frequency of the propagating signal is close to a half of the energy gap in the spectrum of the material.

The prototypical form of the class of models with the localized nonlinearity was proposed in Ref. [30], with the single $\delta$-function multiplying the self-focusing term. While solitons are unstable in the simplest model, they can be readily stabilized by an additional weak periodic linear potential [31]. Solitons can also be made stable in conservative models with a symmetric pair of the $\delta$-functions multiplying the self-focusing cubic nonlinearity, which admit exact analytical solutions for the entire set of symmetric, antisymmetric, and asymmetric modes [32].

While Eq. (1) with the ideal $\delta$-functions is used to obtain analytical solutions for symmetric and antisymmetric patterns, in the numerical simulations the $\delta$-functions are replaced by sufficiently narrow Gaussians,

$$\delta(x) \approx \left(\sigma\sqrt{\pi}\right)^{-1} \exp\left(-x^2/\sigma^2\right), \qquad (2)$$

typically with finite width $\sigma \sim 10^{-2} L$ (all examples of numerical simulations displayed below are produced for $L=1$ and $\sigma = 0.03$). The stability of stationary modes was tested by means of direct simulations using split-step Fourier method. Random perturbation at the 5% amplitude level was added to the initial conditions, in all the cases reported in the paper.

The rest of the article is organized as follows. In Section II, we start by the consideration of the case of the self-focusing nonlinearity, i.e., $B>0$ in Eq. (1), when stable symmetric and antisymmetric patterns actually amount to superpositions of the modes supported by each HS in isolation. Most interesting is the case of the self-defocusing nonlinearity ($B<0$), which gives rise to new nontrivial multi-peak patterns and the multi-stability. It is considered in Section III. Then, in Section IV we address the setting with $B=0$, when the nonlinearity is represented solely by the cubic loss acting at the HSs. In that case, multi-stability is possible too. The paper is concluded in Section V.

## II. SYMMETRIC AND ANTISYMMETRIC MODES IN THE SYSTEM WITH SELF-FOCUSING

### A. The general analysis

As suggested in Ref. [16], exact symmetric and antisymmetric stationary solutions for Eq. (1) with a pair of ideal $\delta$-functions can be looked for in the following forms, respectively:

$$u_{\text{symm}}(x,z) = A_{\text{symm}} \exp\left[\frac{i}{2}\left(Q_r^2 - \frac{\gamma^2}{Q_r^2}\right)z\right]$$

$$\times \begin{cases} \cosh\left[(Q_r - i\gamma/Q_r)x\right]/\cosh\left[(Q_r - i\gamma/Q_r)L\right], & \text{at } |x| < L, \\ \exp\left[-(Q_r - i\gamma/Q_r)(|x| - L)\right], & \text{at } |x| > L; \end{cases} \quad (3)$$

$$u_{\text{antisymm}}(x,z) = A_{\text{antisymm}} \exp\left[\frac{i}{2}\left(Q_r^2 - \frac{\gamma^2}{Q_r^2}\right)z\right]$$

$$\times \begin{cases} \sinh\left[(Q_r - i\gamma/Q_r)x\right]/\sinh\left[(Q_r - i\gamma/Q_r)L\right], & \text{at } |x| < L, \\ \text{sgn}(x) \cdot \exp\left[-(Q_r - i\gamma/Q_r)(|x| - L)\right], & \text{at } |x| > L, \end{cases} \quad (4)$$

with positive real parameter $Q_r$ and real amplitude $A_{\text{symm}}$ or $A_{\text{antisymm}}$. These *ansätze* automatically satisfy the linear equation at $x \neq \pm L$ and the conditions of the continuity of the wave function at the HS points, $x = \pm L$. Integration of Eq. (1) in infinitesimal vicinities of $x = \pm L$ yields a condition for the jumps of the first derivatives at these points. After simple manipulations, the jump condition can be cast into the following form, for Eq. (3) and Eq. (4), respectively:

$$\frac{\gamma/Q_r + iQ_r}{1 + \exp\left[-2L(Q_r - i\gamma/Q_r)\right]} = (\Gamma_1 + i\Gamma_2) + (iB - E)A_{\text{symm}}^2, \quad (5)$$

$$\frac{\gamma/Q_r + iQ_r}{1 - \exp\left[-2L(Q_r - i\gamma/Q_r)\right]} = (\Gamma_1 + i\Gamma_2) + (iB - E)A_{\text{antisymm}}^2. \quad (6)$$

Actually, each complex equation (5) or (6) determines two real positive parameters, *viz.*, $Q_r$ and $A_{\text{symm}}^2$ or $A_{\text{antisymm}}^2$, respectively. For given values of $B$, $E$, $\Gamma_1$, $\Gamma_2$, and $\gamma$, these equations were solved numerically, by splitting them into real and imaginary parts, and selecting physical solutions, with $Q_r > 0$ and $A^2 > 0$.

Generally, Eqs. (5) and (6) give rise to multiple roots for $Q_r$ when $\Gamma_1$ is sufficiently large. Most often, the largest value of $Q_r$ corresponds to the simplest (fundamental) patterns, with two maxima of $|u(x)|^2$ located at or close to $x = \pm L$, see Figs. 1 and 5 below. If the maxima are located at $x = \pm L$, they are sharply peaked (see Fig. 1), due to the jump conditions (5) and (6) at the HSs, in comparison with smooth peaks featured by the solutions corresponding to smaller roots $Q_r$. The smaller roots generate *multi-peak* patterns (instead of the fundamental ones), with additional peaks placed between the basic ones (see Figs. 4, 6, 7, 13 below). As suggested by the character of the (anti)symmetry, the number of the extra peaks generated by exact solutions (3) and (4) is always odd and even for the symmetric and antisymmetric patterns, respectively.

It is relevant to mention that, in the case of the single $\delta$-functional HS, set at $x=0$ [which corresponds to $L\rightarrow\infty$, in terms of Eq. (1)], the exact solution for pinned solitons, in the form of

$$u(x,z) = \sqrt{\frac{Q_r - \Gamma_2}{B}} \exp\left[\frac{i}{2}\left(Q_r^2 - \frac{\gamma^2}{Q_r^2}\right)z - \left(Q_r - i\frac{\gamma}{Q_r}\right)|x|\right]$$

[cf. Eq. (3)], where $Q_r$ is determined by the quadratic equation, $EQ_r^2 - (\Gamma_2 E + \Gamma_1 B)Q_r + \gamma B = 0$, exists for $\Gamma_1\Gamma_2 > \gamma$. Simultaneously, this condition is the instability threshold of the trivial solution, $u=0$, against small perturbations around the HS, i.e., the pinned solitons are generated by the localized instability of the zero state. In the opposite limit of $L\rightarrow 0$, when the two HSs merge into a single one, with double strengths $2\Gamma_1$ and $2\Gamma_2$, the threshold condition takes, accordingly, the form of $4\Gamma_1\Gamma_2 > \gamma$. When $L$ increases from zero to finite values, the threshold value of γ rapidly falls from $4\Gamma_1\Gamma_2$ to $\Gamma_1\Gamma_2$, i.e., the double HS with separation $L \sim 1$ already acts as two individual single HSs, with the threshold value of γ being close to $\Gamma_1\Gamma_2$.

Numerical counterparts of the analytical solutions were found from the integration of the stationary version of Eq. (1), with the $\delta$-functions replaced by approximation (2). The stability of the patterns was identified, as mentioned above, by direct simulations of Eq. (1) with the regularized $\delta$-functions and random perturbations added to the initial conditions.

**B. Symmetric solutions**

Collecting analytical and numerical results obtained with the self-focusing local nonlinearity, $B>0$, we arrive at a simple conclusion: only the fundamental symmetric modes, corresponding to the largest root of $Q_r$ in Eq. (5), may be stable in this case, while all the higher-order multi-peak patterns, which correspond to smaller $Q_r$, are unstable. A typical example of the stable fundamental symmetric mode is shown in Fig. 1.

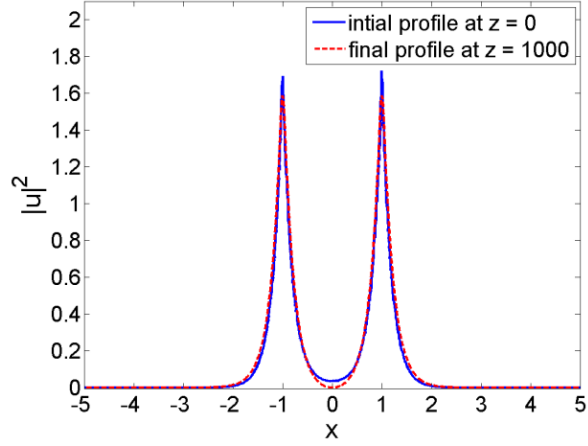

FIG. 1. (Color online) Stable evolution of the symmetric solitary pulse, corresponding to the largest root of Eq. (5), $Q_r = 2.6220$, for $B = E = \Gamma_2 = \gamma = L = 1$, $\Gamma_1 = 2$ in Eq. (1) and $\sigma = 0.03$ in Eq. (2). The blue curve represents analytical solution (3), which was used as the input in the simulations. The red dashed (lower) curve, which attains zero at $x = 0$, depicts the steady-state solution. The discrepancy between the two curves is due to the use of the regularized $\delta$-functions (2) in the numerical simulations.

The stability of the fundamental double-peak modes is mainly controlled by nonlinearity $B$ and local gain $\Gamma_1$, as shown in Fig. 2 (this plot combines the data obtained for $B = 2, 3, 4, 6, 8, 10$). Roughly speaking, the fundamental modes remain stable if the self-focusing is not too strong to cause the collapse, and the rate of the energy pumping by the localized gain is not too high. The value of the linear potential associated with the HS, $\Gamma_2$, is less significant in terms of the stability issue, as long as it remains positive, corresponding to an attractive potential. The fundamental symmetric mode suffers destabilization and transformation into a localized chaotic state for $\Gamma_2 < 0$, which corresponds to a repulsive linear potential, see a typical example in Fig. 3.

It is relevant to compare the stability of the dissipative single-peak solitons, pinned to a single HS, with their symmetric fundamental double-peak counterparts pinned to the dual HSs, with identical parameters for each HS (recall that, as concerns the nonlinearity, in this section we consider the case of self-focusing, $B>0$, and nonzero nonlinear loss, $E > 0$.). The comparison gives rise to the following conclusions. If the linear potential is attractive, $\Gamma_2 > 0$, the stability conditions for the dissipative solitons pinned to the single and double HSs are similar, i.e., the pinned solitons are unstable when the self-focusing is strong and local gain ($\Gamma_1$) is high. On the other hand, when the linear potential is repulsive ($\Gamma_2 < 0$), the instability of the symmetric double-peak solitons sets in at essentially *lower* magnitudes of $B$, $\Gamma_1$ and $|\Gamma_2|$, than it happens for their single-peak counterparts pinned to the single HS. An explanation to the weaker stability of the double-peak pattern in the latter case is that small perturbations, expelled by the repulsive potential from a vicinity of one HS, may hit the other one and get amplified there.

In particular, for $\Gamma_2 > 0$ it is easy to check that the stability border for the fundamental symmetric modes, displayed in Fig. 2, takes nearly the same shape as for the single-peak modes pinned to the single HS, which were introduced in Ref. [16] (actually, such a stability map was not reported in that work). Moreover, the stability boundary is virtually the same for the fundamental antisymmetric modes (see below). These facts imply that, as a matter of fact, the stable symmetric and antisymmetric modes supported by the symmetric set of two HSs, in the case of the self-focusing localized nonlinearity ($B>0$), may be understood as straightforward superpositions of the fundamental single-peak states pinned to each HS separately.

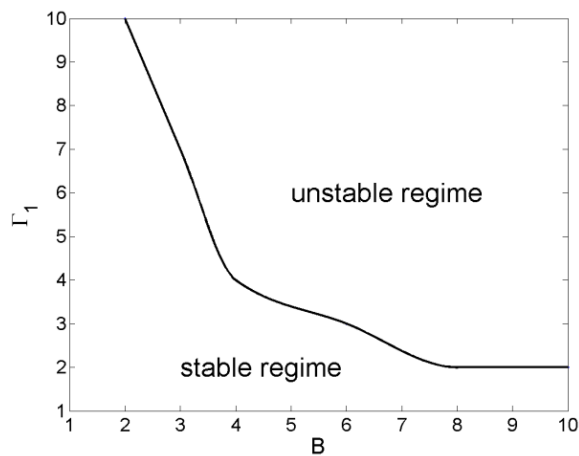

FIG. 2. The stability boundary for the fundamental symmetric and antisymmetric solutions in the plane of the coefficients of the self-focusing nonlinearity ($B$) and linear gain ($\Gamma_1$) acting at each hot spot. Other parameters are $E=1, \Gamma_2=2$.

In the present case of local self-focusing, all the multi-peak (higher-order) symmetric modes, corresponding to smaller roots of Eq. (5) for $Q_r$, are unstable and quickly rearrange themselves into the stable fundamental mode with the same symmetry, corresponding to the largest root $Q_r$ (provided that the latter mode is stable), see a typical example of the rearrangement in Fig. 4. On the other hand, if the fundamental mode is itself unstable, being located above the stability boundary in Fig. 2, its higher-order counterparts transform into chaotic patterns similar to the one displayed in Fig. 3.

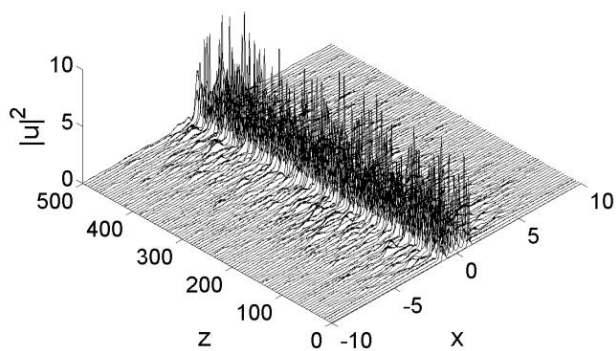

(3) FIG. 3. Unstable evolution of the fundamental symmetric mode for $B=2$, $E=1$, $\Gamma_1=6$, and $\Gamma_2=-2$, which corresponds to the largest root of Eq. (5) at these values of the parameters, $Q_r=9.7958$.

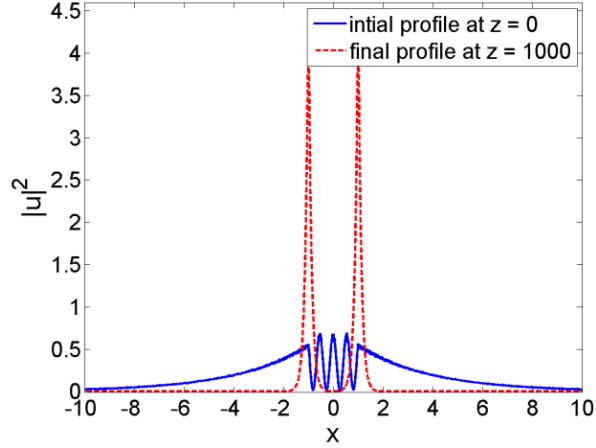

FIG. 4. (Color online) The spontaneous transformation of an unstable multi-peak symmetric pattern, corresponding to a smaller (second) root of Eq. (5), $Q_r = 0.1719$, into the stable fundamental soliton related to the largest root, $Q_r = 4.7915$, for the same parameters, $B = E = \Gamma_2 = 1, \Gamma_1 = 4$, in the case of the local self-attraction.

Finally, in the particular case of the system with local self-focusing and no nonlinear loss ($B > 0, E = 0$), all the stationary patterns (symmetric and antisymmetric ones alike) turn out to be unstable. In this case, the entire system switches into a "turbulent" state, if $\Gamma_2 \geq 0$, or if $\Gamma_2$ is negative and $\Gamma_1$ is large enough. On the other hand, for $\Gamma_1$ being small and $\Gamma_2 < 0$, the wave field in the model with $B > 0, E = 0$ decays to zero (not shown here in detail).

## C. Antisymmetric solutions

In the case of local self-focusing, $B > 0$, general properties of antisymmetric solutions are quite similar to those of their symmetric counterparts. Only the fundamental mode, with two peaks of $|u(x)|^2$ tacked to the HSs, which corresponds to the largest root $Q_r$ of Eq. (6), is stable (inside virtually the same parameter area as in Fig. 2), see an example in Fig. 5. All higher-order antisymmetric patterns, with additional peaks occurring between the HSs, which correspond to smaller roots $Q_r$, are unstable, spontaneously rearranging themselves into the fundamental mode, as shown in Fig. 6. In case the fundamental antisymmetric mode is unstable, slightly perturbed higher-order ones will evolve into chaotic patterns (not shown here in detail).

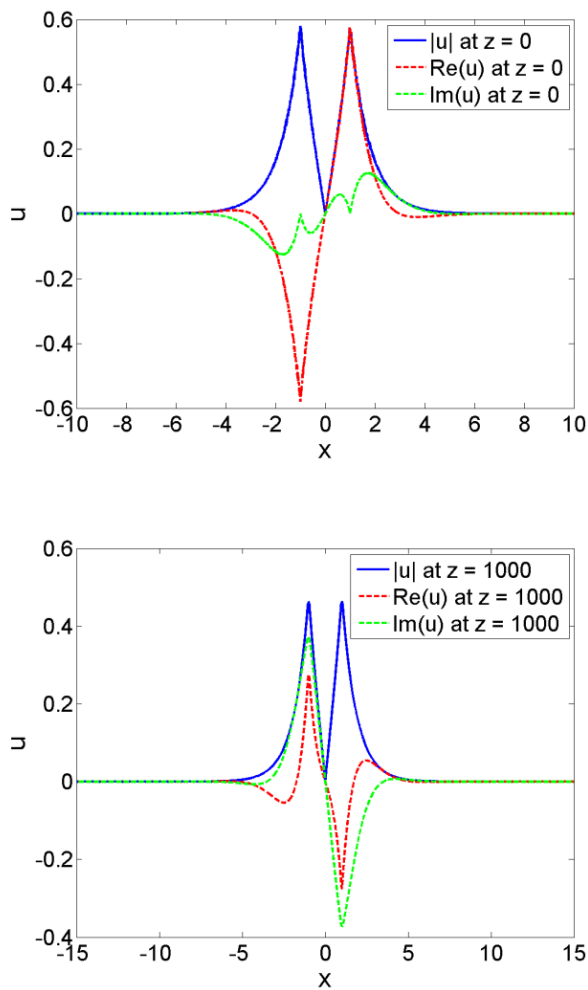

FIG. 5. (Color online) The stable evolution of a fundamental antisymmetric mode, which corresponds to the largest root of Eq. (6), $Q_r = 1.2589$, with constants $B = E = \Gamma_1 = \Gamma_2 = 1$. The relation between the real and imaginary parts of the wave field changes due to a phase shift between the initial and final states.

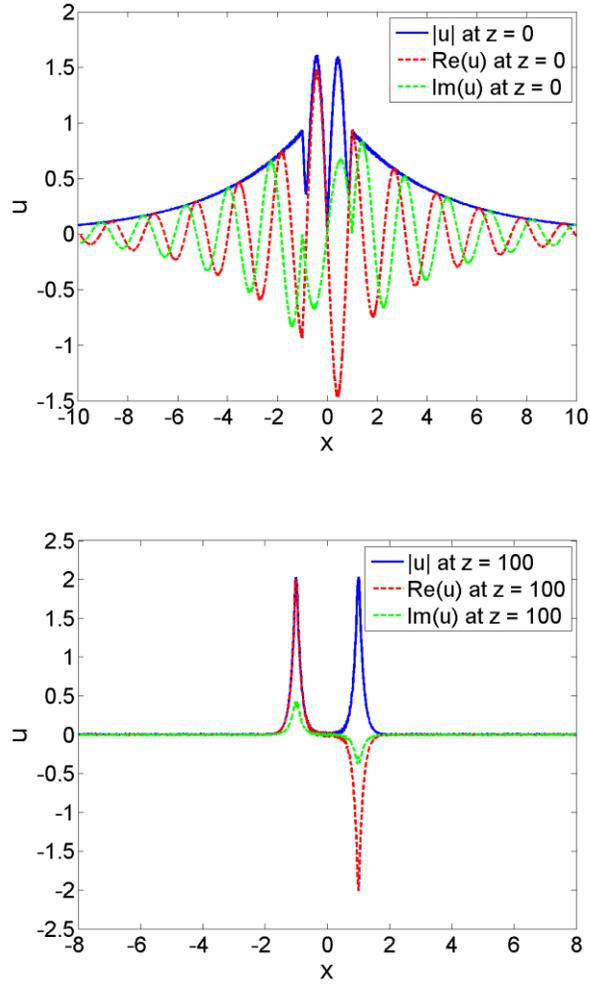

FIG. 6. (Color online) Spontaneous transformation of an unstable higher-order (multi-peak) antisymmetric pattern, corresponding to a smaller root, $Q_r = 0.2712$ (the second root, there being only two roots in this case), of Eq. (6) with $E = \Gamma_2 = 1, \Gamma_1 = 4, B = 2$, into its stable counterpart corresponding to the largest root, $Q_r = 8.7720$.

# III. SYMMETRIC AND ANTISYMMETRIC PATTERNS IN THE SYSTEM WITH SELF-DEFOCUSING

## A. The system including the local nonlinear loss ($E > 0$)

### 1. The meaning of the multi-stability

It was concluded above that, in the case of the self-focusing local nonlinearity ($B > 0$), all the stable symmetric and antisymmetric modes supported by the HS pair amount to straightforward superpositions of stable single-pinned states separately pinned to each HS. The situation becomes essentially more interesting in the case of the self-defocusing nonlinearity ($B < 0$), while the nonlinear loss may or may not be present ($E > 0$ or $E = 0$, respectively). We first consider the generic case of the system including the nonlinear loss.

In this case, the exact symmetric and antisymmetric solutions, based on Eqs. (3) and (4), respectively, may be *stable* not only for the largest roots $Q_r$ of Eqs. (5) or (6), but also for some smaller ones. In other words, higher-order wave profiles with additional peaks inserted between the HSs may be stable too. A drastic expansion of the manifold of stable patterns is thus possible. Such a scenario, where two or more stable profiles coexist for a set or parameter values, is termed "multi-stability" in this paper.

Numerical simulations show that those profiles which are, nevertheless, unstable quickly evolve into the solutions corresponding to larger roots of $Q_r$. As a simple example, the symmetric solutions obtained for $E = 1, B = -1, \Gamma_1 = 6, \Gamma_2 = 2$, with three and five peaks, corresponding to roots $Q_r = 0.2614$ and $Q_r = 0.1562$ of Eq. (5), respectively, are *stable*, while an additional seven-peak mode, which corresponds to $Q_r = 0.1087$, is unstable and evolves into the three-peak state.

The salient feature reported in the previous section for the case of the self-focusing was that only the fundamental two-peak mode might be stable, while higher-order profiles with additional peaks placed between the two HSs rapidly relaxed towards the fundamental mode. In the self-defocusing regime, surprisingly, the solution for the fundamental two-peak mode does *not* exist in parts of the parameter space. Specifically, when $B < 0$, $E \geq 0$, or $B = 0$, $\Gamma_2 < 0$, the fundamental two-peak mode is absent in some parametric regions.

*2. Symmetric modes*

For symmetric solutions in the model with the self-defocusing, it has been found that, if the fundamental double-peak mode exists, then it represents the *only* stable solution, whilst all the higher-order modes are unstable, spontaneously transforming into the fundamental one. This situation is thus similar to that for the symmetric modes under the self-focusing, as reported in the previous section.

The multi-stability arises if the fundamental two-peak mode is *absent*. While it is generally true that most of the multi-peak modes which are unstable rearrange themselves into stable states corresponding to larger roots of $Q_r$, in some cases the existing solutions corresponding to the largest roots (which are *not* the fundamental modes) may be unstable too. These unstable solutions will then spontaneously evolve into stable modes corresponding to *smaller* roots $Q_r$.

Some of the higher-order solutions predicted by exact wave forms (3) and (4) turn out to be very close to their numerically found stable counterparts, featuring only small local deviations in the structure of the solutions due to the replacement of the ideal $\delta$-functions by the Gaussian approximation (2) in the simulations. On the other hand, the use of Eq. (2) to approximate the $\delta$-functions sometimes gives rise to additional stable numerical solutions, which are attained through the evolution of perturbed analytical wave forms. An extraordinary feature of these additional modes is that they exhibit the number of intermediate peaks between the HSs which may be different, as concerns its parity, from the prediction of the analytical solutions (recall the latter always predicts an odd/even number of extra peaks for the symmetric/antisymmetric solutions, respectively).

The multi-stability of the modes generated by the symmetric analytical input corresponding to Eq. (3) is illustrated by a set of *five* stable coexisting multi-peak modes shown in Fig. 7. In particular, the last panel of the figure displays an essential difference between the established mode and the input wave form, with the change of the size and *parity* of the set of intermediate peaks between the HSs, while a 9-internal-peak mode evolves into a 6-internal-peak one.

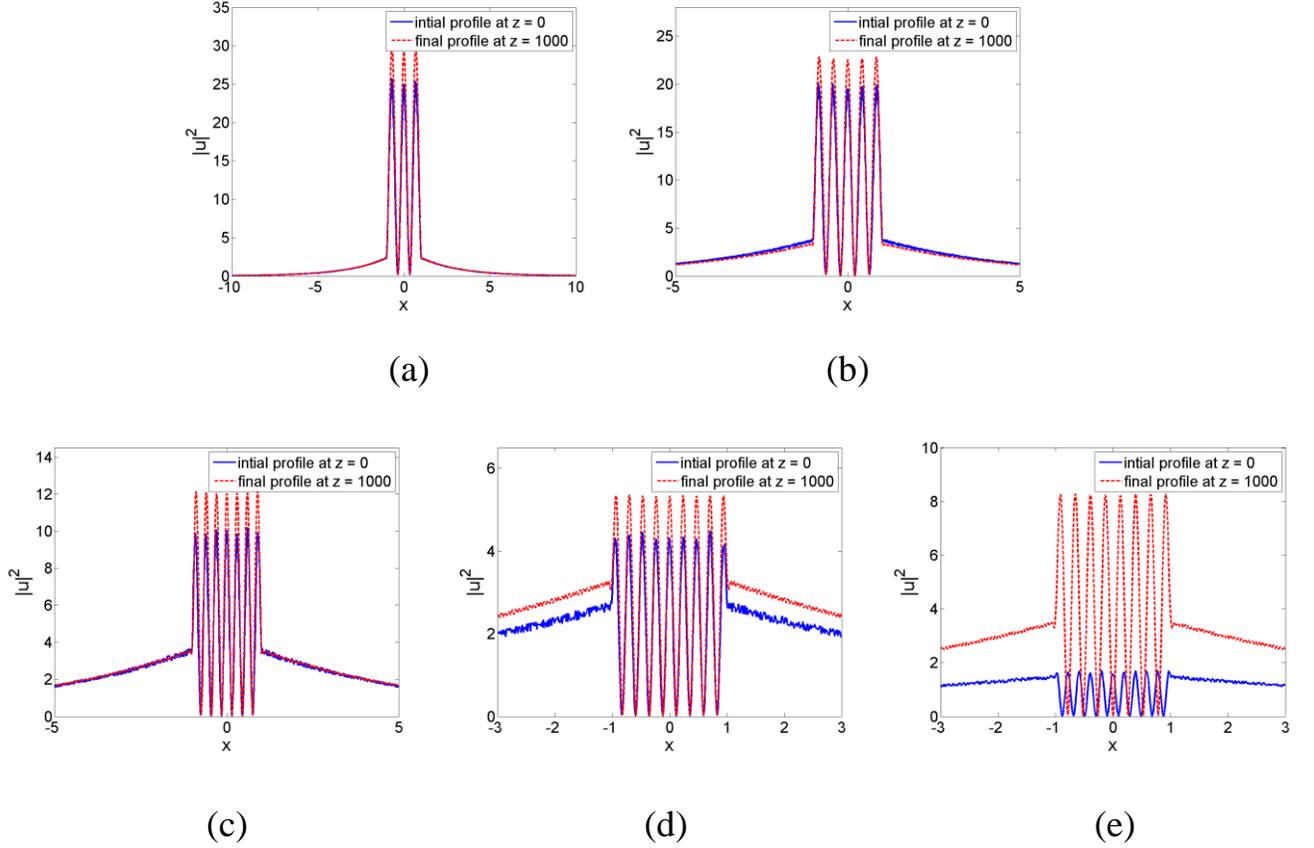

(a) (b)

(c) (d) (e)

FIG. 7. (Color online) A set of multistable multi-peak modes in the case of the self-defocusing local nonlinearity, for $E=1, B=-2, \Gamma_1=10, \Gamma_2=0$. The five modes are generated, respectively, by the symmetric input taken in the analytical form (3), with the following roots of Eq. (5): (a) $Q_r$ = 0.2226; (b) 0.1348; (c) 0.0965; (d) 0.0756; (e) 0.0623. The smallest root, $Q_r$ = 0.0530, corresponds to the unstable mode, which spontaneously evolves into its stable counterpart related pertaining to $Q_r$ = 0.0965 (the evolution is not shown here).

*3. Dependence on the strength of the linear potential* ($\Gamma_2$)

The picture outlined above (the multi-stability of multi-peak patterns) is essentially the same for both signs of $\Gamma_2$, i.e., for both attractive and repulsive linear potentials induced by each HS. Nevertheless, one noteworthy difference in the case of the repulsive potential, $\Gamma_2 < 0$, is that, in addition to the multistable set of stationary modes, a localized chaotically oscillating state may coexist with them, provided that $\Gamma_1$ and $|B|$ are large enough. An example of such an additional robust chaotic mode, which exists in addition to the multi-stability of the stationary states, is displayed in Fig. 8.

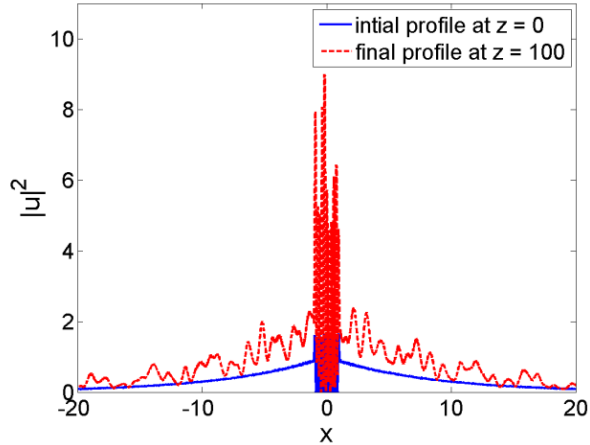

FIG. 8. (Color online) A localized chaotic mode coexisting with four stable multi-peak symmetric patterns in the case of $\Gamma_2 = -4$ and $E = 1, B = -4, \Gamma_1 = 10$. This mode is generated by the symmetric analytical wave form (3) with root $Q_r = 0.0608$ of Eq. (5). At the same values of the parameters, three larger roots, $Q_r = 0.1325; 0.0945; 0.0738$, give rise to a multistable set of stationary multi-peak patterns (not shown in this figure). The largest root, $Q_r = 0.2189$, gives rise to an *unstable* mode, which spontaneously evolves into the stable mode corresponding to $Q_r = 0.0945$ (not shown here either).

The results concerning the multi-stability of modes generated by symmetric initial conditions (3) in the system with self-defocusing nonlinearity are summarized in Fig. 9, in the form of a table which reports the number of stable coexisting stationary modes, together with the number of roots of Eq. (5), i.e., the number of localized symmetric modes predicted by the analytical solution based on the ideal $\delta$-functions. These results include the stable patterns with the "wrong" parity regarding the number of intermediate peaks, but do not include localized chaotic states, such as the one shown in Fig. 8. The increase of local gain $\Gamma_1$ leads to an increase of the number of the coexisting stable modes. Additional numerical data demonstrate that the increase in the strength of the localized self-defocusing, $|B|$, causes an increase or decrease of the number of the stable modes in the cases of the attractive or repulsive linear potential, i.e., $\Gamma_2 > 0$ and $\Gamma_2 < 0$, respectively.

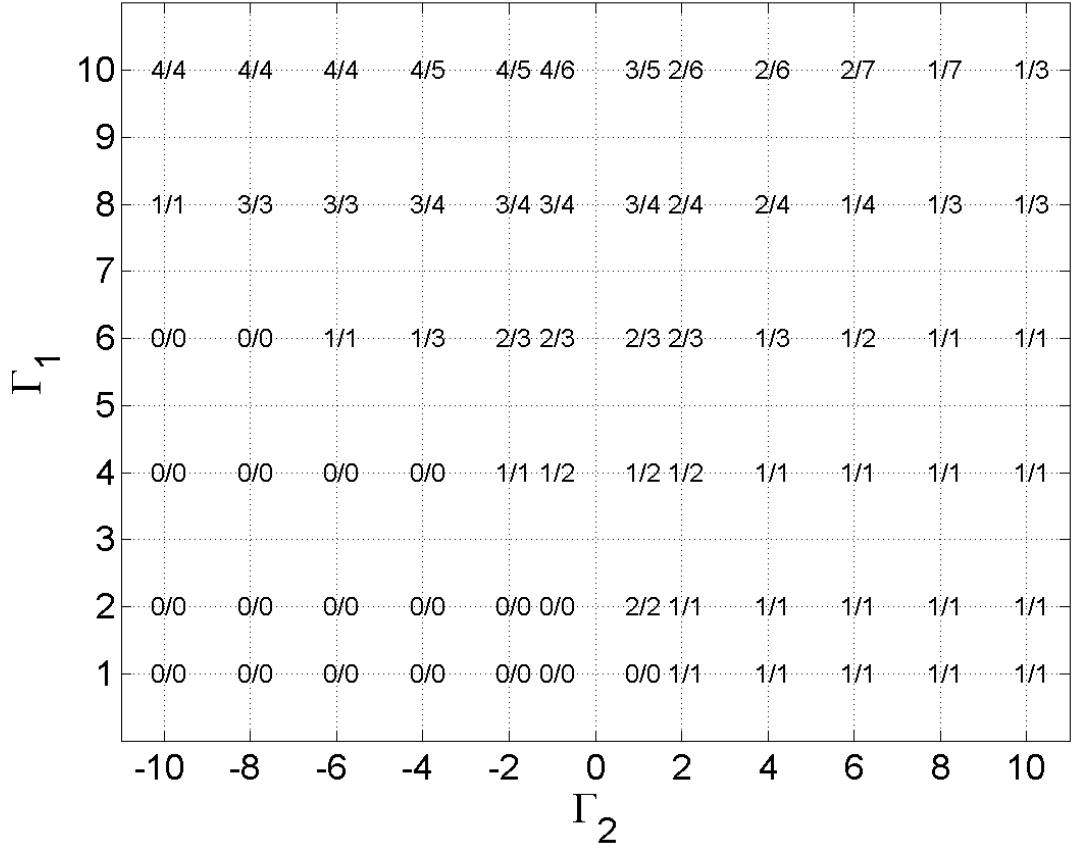

FIG. 9. The multi-stability chart for the modes generated by the symmetric input (3) in the system with the local self-defocusing nonlinearity, in the plane of the parameters characterizing the local gain and linear potential, $\Gamma_1$ and $\Gamma_2$ (recall $\Gamma_2 > 0$ corresponds to the attractive potential). Other parameters are $E = 1$ and $B = -1$. In each entry, the numerator and denominator give the number of the actually existing stable stationary modes and the corresponding number of roots $Q_r$ of Eq. (5). Recall that the fundamental (two-peak) modes *do not exist*, i.e., they do not contribute to the count of the coexisting modes, when the multi-stability occurs in the self-defocusing system.

*4. Antisymmetric modes*

In contrast with the situation for symmetric states outlined above, the fundamental two-peak modes and their higher-order counterparts can coexist as stable entities in the class of antisymmetric solutions. In addition, peaks of those fundamental antisymmetric modes which coexist with their higher-order counterparts, in the case of the multi-stability, are shifted inward, rather than being located exactly at the HSs. We stress that the shift is a property of the exact analytical solution, as given by Eqs. (4) and (6), and is not a result of the replacement of the ideal $\delta$ -functions by their regularized counterparts (2). The shift effect is only observed in the parametric regions of $B < 0$, $E > 0$, or $B = 0$, $\Gamma_2 < 0$, where the multi-stability may occur. The fundamental symmetric mode never features the shift. A typical example of such peak-shifted fundamental antisymmetric mode is displayed in Fig. 10. The fundamental antisymmetric modes with or without the shift of the peaks can exist as stable solutions in the parametric regions of $B < 0$, $E > 0$, or $B = 0$, $\Gamma_2 < 0$.

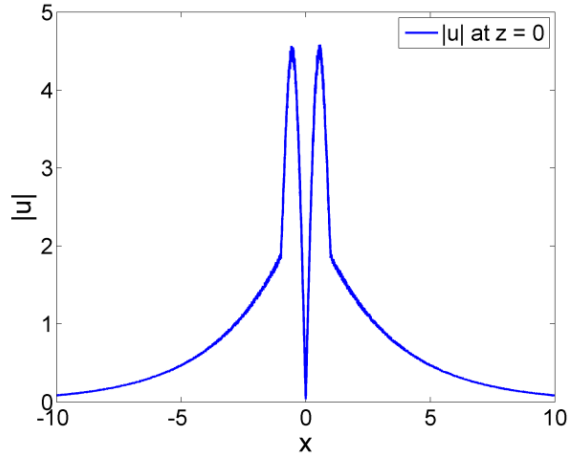

FIG. 10. (Color online) An example of a fundamental antisymmetric mode with its peaks shifted inwards, from $|x| = 1$ to $|x| = 0.545$. The mode corresponds to the largest root of Eq. (6), $Q_r = 0.3447$, for $B = -1$, $E = 1$, $\Gamma_1 = 8$, $\Gamma_2 = 2$. This fundamental mode, together with its two higher-order counterparts, pertaining to $Q_r = 0.1830$ and $Q_r = 0.1221$ (not shown in the figure), are stable. An additional unstable higher-order mode, corresponding to $Q_r = 0.3183$, spontaneously evolves into the fundamental one shown here. The remaining two unstable higher-order modes, corresponding to $Q_r = 0.0909$ and $Q_r = 0.0718$, evolve into the above-mentioned stable solution which pertains to $Q_r = 0.1830$.

In other respects, the study of modes generated by the antisymmetric initial conditions (4) yields results which are similar to those outlined above for the symmetric input (3). The multistability chart for this class of solutions is presented in Fig. 11.

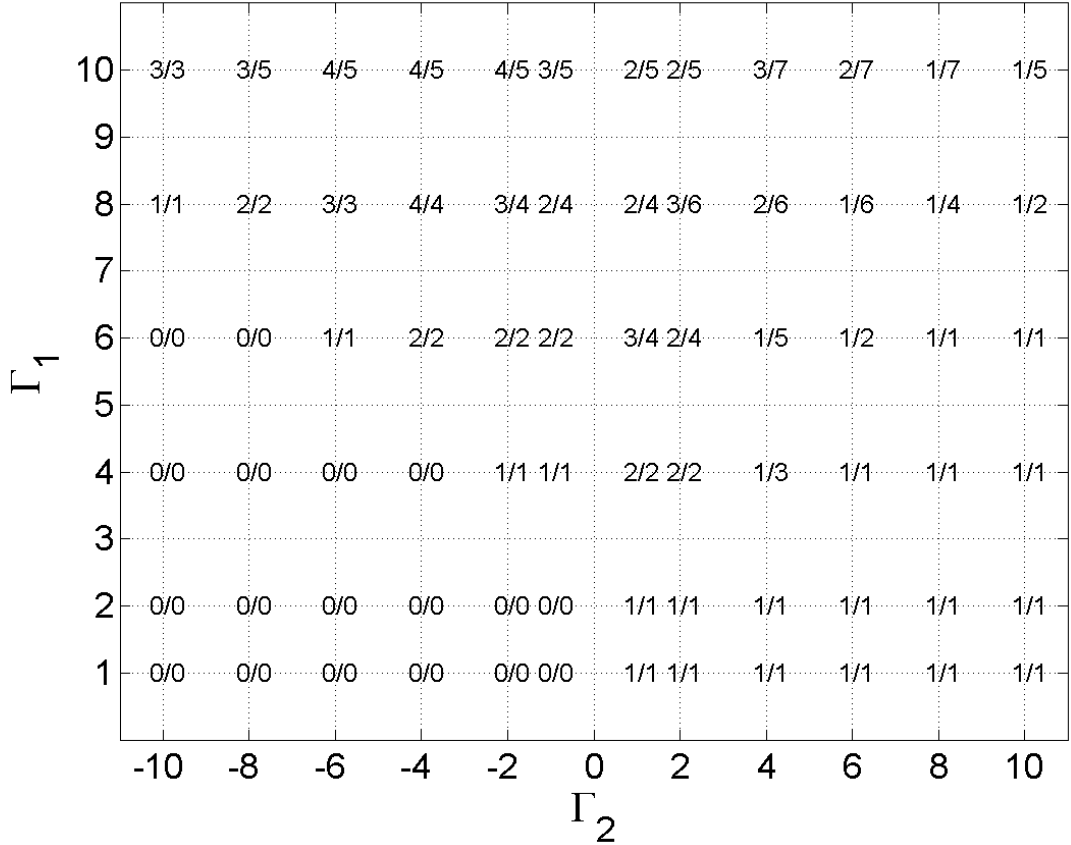

FIG. 11. The same as in Fig. 9, but for modes generated from antisymmetric input (4). In this case, the fundamental two-peak modes coexist with the higher-order ones (i.e., the total number of the coexisting modes includes the fundamental one), in contrast to the situation with the symmetric solutions.

**B. The system without the nonlinear loss ($E = 0$)**

When nonlinearity is represented solely by the local self-defocusing term and nonlinear loss is absent, the system is more sensitive to the deviation of approximation (2) from the ideal $\delta$-function. As a result, the inputs corresponding to analytical solutions (3) and (4) readily generate stable stationary patterns which feature essential differences from the analytical wave forms. In some cases (in fact, for large values of the localized linear gain, $\Gamma_1 \geq 8$, while the corresponding linear potential may be both attractive and repulsive, $\Gamma_2 > 0$ or $\Gamma_2 < 0$), these inputs can also give rise to localized chaotically oscillating modes coexisting with the stable stationary states. The multi-stability chart for the system with $E = 0$, generated by the symmetric inputs (3), is presented in Fig. 12. In the case of $E = 0$, similar to $E > 0$, the symmetric fundamental mode is absent when the multi-stability is observed.

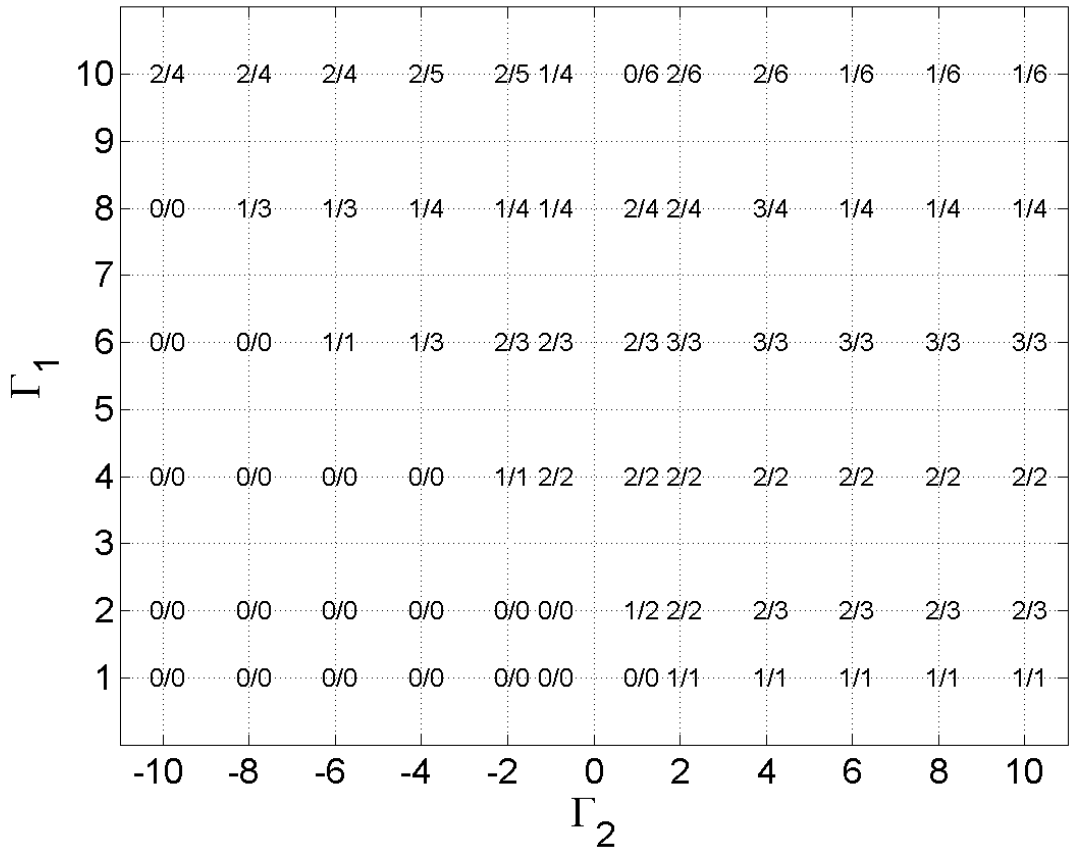

FIG. 12. The same as in Fig. 9, but for the symmetric modes in the system with $E = 0$ and $B = -1$ (actually, $|B| = 1$ may always be fixed by rescaling, in the absence of the nonlinear loss).

The picture generated by the antisymmetric inputs (4) in the case of $E=0$ (see Fig. 13) displays properties similar to those outlined above for the symmetric inputs. Still, a noteworthy difference is that the antisymmetric input makes the self-trapping of stable modes harder to occur for large values of the linear gain, $\Gamma_1$. This difference results in the set of zeros in the numerator of entries in the upper row of the table in Fig. 13, cf. Figs. 9, 11, 12 (no stable modes occur at the respective values of the parameters; in fact, localized chaotically oscillating patterns are found instead).

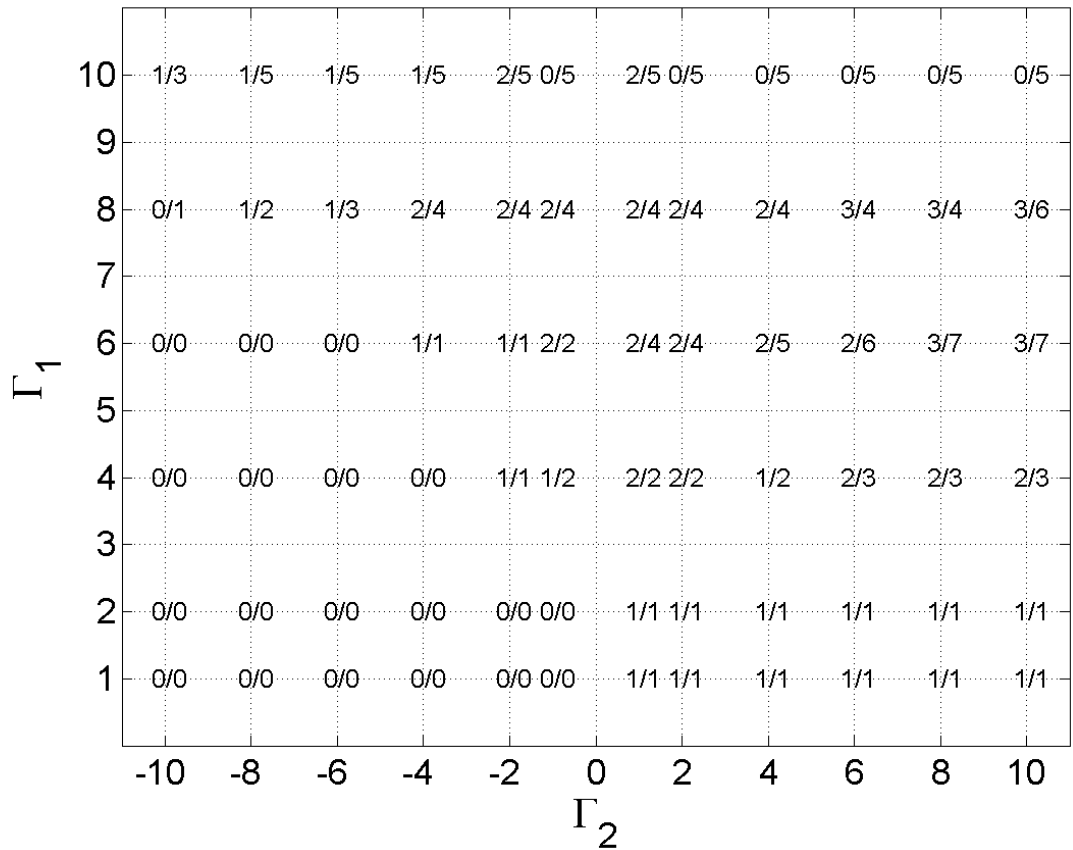

FIG.13. The same as in Fig. 12, but for stationary modes generated by antisymmetric input (4).

## IV. THE SYSTEM WITHOUT THE LOCAL SELF-INTERACTION ($B = 0$)

In the two previous sections, it was demonstrated that the ability of the symmetric HS pair to support stable localized patterns crucially depends on the sign of the self-interaction ($B$), the multi-stability being possible only in the case of self-defocusing, $B<0$. Therefore, it is interesting to study the borderline case of $B=0$, when the nonlinearity is represented solely by the local cubic loss, $E>0$.

The numerical investigation of the system with $B=0$ demonstrates the crucial role of the sign of the linear potential. In the case of the attractive potential, $\Gamma_2 > 0$, the fundamental symmetric modes, which are generated by input (3), taken with the largest root $Q_r$ of Eq. (5), are always stable, while all the higher-order (multi-peak) symmetric modes, generated by smaller roots $Q_r$, are unstable, spontaneously evolving into the fundamental state corresponding to the largest $Q_r$. In this respect, the situation is quite similar to that outlined in Section II for the case of $B>0$. Nevertheless, a difference from that case is observed if $\Gamma_1$ (the local gain) is large enough, while $\Gamma_2$ is sufficiently small, namely, $\Gamma_1 \geq 6$, $0 < \Gamma_2 \leq 1$: As shown in Fig. 14, in addition to the stable fundamental symmetric mode, the simulations of the evolution of symmetric input (3) corresponding to one of smaller roots end up with the establishment of a stable localized breather, which exhibits quasi-regular oscillations (rather than apparently chaotic oscillations in the above-mentioned nonstationary modes, cf. Fig. 8). The stable breather coexists with the stationary fundamental mode. A completely similar picture (not shown here in detail) is observed for the modes generated by antisymmetric wave forms (4).

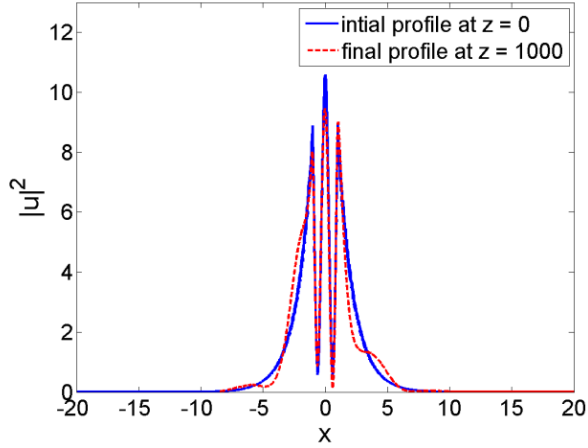

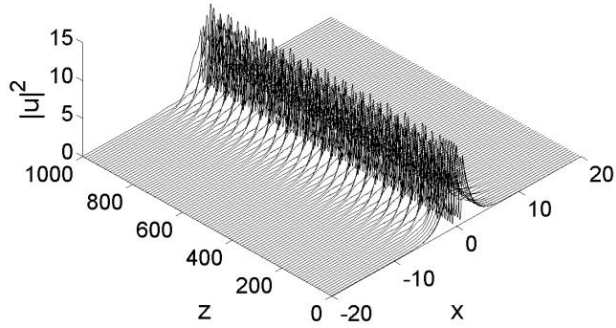

FIG. 14. (Color online) The two panels demonstrate the self-trapping of a stable breather from the symmetric input (3) with $B = 0$, $E = \Gamma_1 = 10$, $\Gamma_2 = 1$, the corresponding root of Eq. (5) being $Q_r = 0.3935$.

The situation turns out to be drastically different for the repulsive linear potential, $\Gamma_2 < 0$, being similar to what was described in the previous section for the self-defocusing nonlinearity, $B < 0$. In this case, the multi-stability is observed, see the corresponding chart in Fig. 15. As before, stable higher-order modes are generated by the symmetric wave forms (3) with $Q_r$ taken as roots of Eq. (5) different from the largest one. Established profiles of some of these modes, produced by the numerical solution, may differ from those predicted by the analytical solution obtained with the ideal $\delta$-functions. At $\Gamma_1 \geq 6$, symmetric inputs (3), with roots $Q_r$ smaller than the largest one, may evolve into localized chaotically oscillating modes (not shown here in detail).

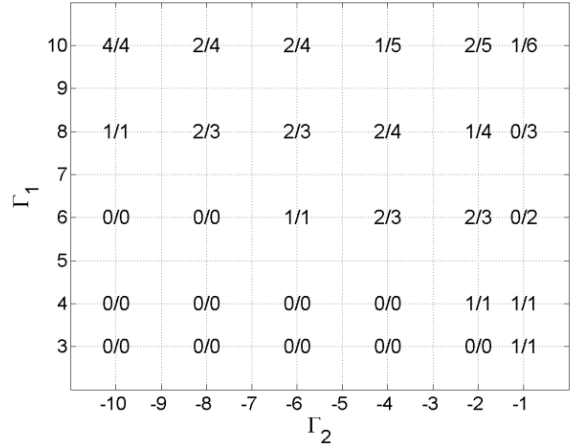

FIG. 15. The same as in Figs. 9 and 12, but for the symmetric modes in the system with $B=0$ and $E=1$ (the strength of the nonlinear local loss may always be scaled to $E=1$ in this case).

Finally, the system demonstrates a similar behavior for $\Gamma_2 < 0$ if antisymmetric input (4) is used, with $Q_r$ taken as roots of Eq. (6). The systematic simulations reveal the multi-stability in this case too, as shown by the chart displayed in Fig. 16.

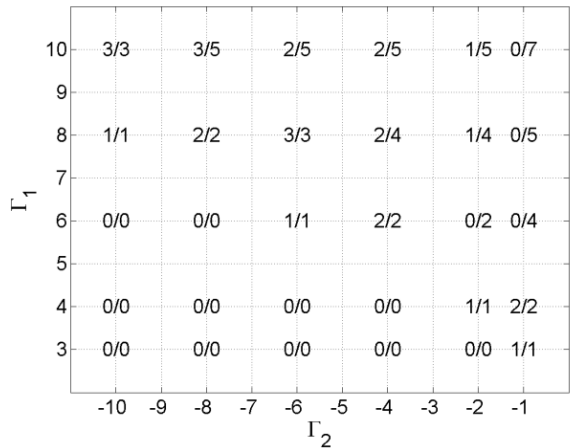

FIG. 16. The same as in Fig. 15, but for stationary modes generated by antisymmetric initial wave form (4).

## V. CONCLUSIONS

Our objective was to analyze the general features of one-dimensional pattern formation in a system where localized modes are supported by a symmetric set of two narrow HSs ("hot spots") carrying the linear gain, linear attractive or repulsive potential, nonlinear self-interaction, and cubic loss. The host medium accounts for the paraxial diffraction and uniform linear loss. This system can be realized in optics, and in Bose-Einstein condensate (BEC) of quasi-particles in solid-state media. In a recent work [16], it was demonstrated that the model with a single HS yields exact solutions, of the generic type, for dissipative solitons pinned to the HS. In this work, we demonstrate that the system with the dual HS opens a way to find a class of exact coexisting multi-peak modes, which may be symmetric or antisymmetric with respect to the underlying HS pair. The most important part of the analysis is the study of the stability of these modes, which was performed by means of systematic direct simulations, with the ideal $\delta$-functions replaced by regularized expressions (2). The analysis has demonstrated the crucial role of the sign of the self-interaction tacked to the HSs: only the fundamental modes, featuring two peaks of the local power at the HSs and no additional peaks between them, may be stable in the case of the self-focusing [ $B > 0$ in Eq. (1)], while all the higher-order multi-peak states are unstable, spontaneously evolving into the fundamental one (unless it is itself unstable). Actually, the fundamental modes, both symmetric and antisymmetric ones, can be understood (in the case of $B > 0$) as superpositions of single-peak dissipative solitons separately pinned to each HS. In particular, the stability domain for the fundamental two-peak modes is practically identical to that for the individual single-peak solitons, provided that the linear potential of the HSs is attractive, $\Gamma_2 > 0$ (that domain, presented in Fig. 2 here, was not reported in previous work [16]). The fundamental double-peak solitons are unstable in the parameter regime of $B > 0$ and $\Gamma_2 < 0$ (with the repulsive linear potential).

The situation turns out to be essentially more interesting for the self-defocusing localized nonlinearity, i.e., $B < 0$ in Eq. (1). In this case, it has been found that the double HS gives rise to a complex multi-stability, with up to *eight* stably coexisting multi-peak patterns (counting both symmetric and antisymmetric ones, see Figs. 9 and 11); in addition, they may coexist with effectively stable localized modes featuring chaotic oscillations. Thus, the present model produces a set of multi-stable multi-peak patterns in the exact analytical form. A remarkable peculiarity is that the fundamental double-peak symmetric mode does not exist in the case of the multi-stability of the symmetric solutions; on the other hand, the existence of the fundamental antisymmetric two-peak mode is compatible with the multi-stability of the anti-symmetric solutions.

In the intermediate case of a system without the localized self-interaction [$B=0$ in Eq. (1)], when the nonlinearity is represented solely by the local cubic loss, the situation is similar to that observed in the cases of $B>0$ and $B<0$, if the linear potential is attractive ($\Gamma_2 > 0$) or repulsive ($\Gamma_2 < 0$), respectively. A noteworthy feature of the case with $B=0$ and $\Gamma_2 > 0$ is that the stable fundamental symmetric and antisymmetric modes can coexist with a localized breather featuring quasi-regular oscillations.

Unlike some models with a smooth spatial modulation of the local gain [20-22], the present system, which is based on the pair of ideal $\delta$-functions or their regularized counterparts (2), does not reveal asymmetric modes, anywhere in the explored parameter space. Therefore, it may be interesting to find a border between the gain-modulation landscapes which do or do not give rise to stable asymmetric states. Another relevant extension may be the analysis of the dual-HS setting in the two-dimensional geometry, although analytical solutions are not likely to be found in that case (a conservative counterpart of such a two-dimensional model, with a set of two nonlinear circles embedded into a linear host medium, was recently studied in Ref. [33]).

**ACKNOWLEDGEMENT**

Partial financial support to this work has been provided by contract HKU7120/08E from the Research Grants Council of Hong Kong.